# UV- and X-ray-activated broadband NIR garnet-type Ca$_3$Ga$_2$Sn$_3$O$_{12}$:Fe$^{3+}$ phosphors with efficient persistent luminescence

Yongjie Wang[*ab], Qingzhou Bu[a], Dongshuo Li[a], Shuai Yang[a], Li Li[a], Guotao Xiang[a], Sha Jiang[a], Ying Chang[a], Chuan Jing[a], Xianju Zhou[*ab], Lev-Ivan Bulyk[c] and Andrzej Suchocki[*c]

[a]College of Sciences, Chongqing University of Posts and Telecommunications, Chongqing 400065, China

[b]Chongqing Key Laboratory of Photoelectronic Information Sensing and Transmitting Technology, Chongqing University of Posts and Telecommunications, Chongqing 400065, China

[c]Institute of Physics, Polish Academy of Sciences, Al. Lotników 32/46, 02-668 Warsaw, Poland

**Abstract**

Near-infrared phosphor-converted light-emitting diodes (NIR pc-LEDs) are compact light sources of great interest for NIR spectroscopy applications. Beyond typical Cr$^{3+}$-activated NIR-emitting phosphors, there exists a strong demand for Cr$^{3+}$-free alternatives with high efficiency and broadband emission to rich the landscape of NIR luminescent materials and extend their range of application fields. Here, we report a series of Fe$^{3+}$-activated Ca$_3$Ga$_2$Sn$_3$O$_{12}$ garnet-type phosphors exhibiting broadband NIR emission in the 650-1000 nm range attributed to $^4T_1(G) \rightarrow {}^6A_1(S)$ transition, with a maximum at 754 nm and a FWHM of 89 nm upon UV excitation. The spectroscopic results were analyzed according to the Tanabe-Sugano theory from which the crystal field parameter $D_q$ and Racah parameters $B$ and $C$ were obtained for the octahedrally coordinated Fe$^{3+}$ ion. Notably, the NIR persistent luminescence lasting over 1 h was detected following UV or X-ray irradiation. The possible mechanism involving electron traps was proposed to explain the observed persistent luminescence. Furthermore, a NIR pc-LED was fabricated by coating synthesized phosphor on a UV chip, and its performance was evaluated to assess its potential suitability as a NIR light source. Our discovery of novel type of nontoxic $Fe^{3+}$-activated broadband NIR luminescence phosphors with efficient NIR persistent luminescence paves the way for discovering $Cr^{3+}$-free multifunctional NIR luminescence materials, thereby expanding their application possibilities.

1. Introduction

Near-infrared (NIR) spectroscopy offers several advantages over traditional techniques, including rapid detection, nondestructive and noninvasive examinations and non-contact capabilities [1],[2],[3],[4]. Conventional high-performance spectrometers are typically based on bulky and costly systems with large dispersive components, long optical path length, and movable mechanisms. To meet emergent requirements of various application scenarios where the portability, cost-effectiveness, intelligence, compactness and power consumption are crucial parameters, such as portable portability or mobile equipment, wearable sensing devices (e.g., for health monitoring[5], food safety analyzing [6] and biomedicine detecting [7]) and other micromachined components [8],[9],[10] create a great synergy with the inherent advantages of NIR spectroscopy during the past decades. However, the ongoing pursuit of miniaturization and intelligence in integrated spectrometers faces challenges due to the intrinsic limitations of traditional NIR light sources, such as tungsten halogen lamps, globars, and tunable laser, which are characterized by large volumes, high energy consumption and low efficiency. An alternative candidate for NIR light sources, currently under development, is NIR phosphor-converted light-emitting diode (NIR pc-LEDs) [11],[12],[13]. These devices are fabricated by integrating broadband NIR phosphors with a LED chip for excitation. NIR pc-LED offers remarkable advantages over traditional light sources, namely the smaller size, spectral stability, cost-effectiveness and customized tunable broad spectral distribution. Hence, the design and development of promising broadband NIR phosphors hold significant importance in optimizing the performance of NIR pc-LED and extending the scope of applications for smart and miniaturized NIR spectroscopy devices.

$Cr^{3+}$ ([Ar]$3d^3$), as an ideal activator for generating broadband NIR emission when

experience a weak octahedral crystal field, have been widely investigated in garnet [14],[15],[16], langasite[17],[18],[19], perovskite [20],[21], borophosphate [22],[23] and spinel [24], pyroxene [25],[26] to name a few. The leading materials of $Cr^{3+}$-activated broadband NIR phosphors are garnet-type phosphors, which display some advantages over other materials, such as high internal quantum efficiency (IQE) and good thermal stability **Błąd! Nie zdefiniowano zakładki.**[27],[28],[29]. However, the application of Chromium ions may result in an increased risk of toxicity in healthy individuals [30], especially with the oxidation of $Cr^{3+}$ to $Cr^{6+}$, thereby restricting their practical applications in certain fields, particularly in long-term in vivo applications. Consequently, there is an urgent requirement to explore environmental-friendly Chromium-free NIR-emitting phosphors suitable for both in vitro and vivo applications. Recently, $Sb^{3+}$ [31],[32], $Mn^{2+}$[33], $Fe^{3+}$ [34],[35] and $Eu^{2+}$ [36],[37],[38] activated broadband NIR phosphors have been reported. Among these, $Fe^{3+}$ ion ($[Ar3d^5]$) stands out as an essential element in numerous biological systems, being non-toxic element and regarded as a favorable dopant. Typically, $Fe^{3+}$ tends to produce broadband emission in red or NIR range when occupying tetrahedral sites. The resulting wavelength can be further extended when $Fe^{3+}$ ions are situated in octahedral sites, primarily due to an amplified crystal field splitting effect. However, achieving user-designed $Fe^{3+}$-activated broadband NIR phosphors with high efficiency becomes challenging in an octahedral environment, where the *d-d* transitions of $Fe^{3+}$ are notably weaker due to the stricter adherence to Laporte selection rules compared to tetrahedral coordination.

In this study, we present $Fe^{3+}$-activated garnet-type $Ca_3Ga_2Sn_3O_{12}$ (CGSG:Fe) phosphors. Within CGSG structure, $Fe^{3+}$ ion occupies $Ga^{3+}$ octahedral site (with $C_{3i}$ point symmetry) which undergoes weak trigonal distortion and thus exhibits a strong broadband NIR emission in 650-1000 nm range, with a maximum around 754 nm and a FWHM of 89 nm. The crystal-field strength, Racah parameters and electron-phonon coupling have been analyzed based on spectroscopic results. Additionally, the NIR persistent luminescence lasting for 1 h was observed after UV or X-ray irradiation. The mechanism for observed persistent luminescence was proposed. Furthermore, the NIR pc-LED was fabricated by coating the synthesized phosphor onto a UV chip, and its potential as a compact light source was examined. The results highlight the potential application of $Fe^{3+}$-activated $Ca_3Ga_2Sn_3O_{12}$ (CGSG:$Fe^{3+}$) phosphors for NIR light source and offer promising avenues for the design of $Cr^{3+}$-free multifunctional NIR luminescence materials.

## 2. Experimental section

### 2.1 Synthesis of Materials

The $Ca_3Ga_{2-x}Fe_xGe_3O_{12}$ (x=0-0.04) garnet-type structure phosphors, here abbreviated as CGSG:xFe, were synthesized using the high-temperature solid-state reaction method. All chemicals employed were of analytical grade and used as received without further purification. In the process of preparation, $Ca_2O_3$ (99.9%, aladdin, Shanghai, China), $Ga_2O_3$ (99.99%, aladdin), $GeO_2$ (99.99%, aladdin), and $Fe_2O_3$ (99.99%, aladdin) were used as starting materials in the appropriate stoichiometric ratios. Each mixture was dispersed in a small amount of ethanol and adequately ground for 20 min in an agate mortar. Subsequently, the mixtures were transferred to open alumina crucibles and initially calcined at 1173 K for 6 h in an air atmosphere to decompose the carbonates. After this initial firing, the resulting products were ground again and then subjected to calcination at 1473 K for 40 h with an intermediate grounding. After the samples were cooled to room temperature, they were ground into fine powders for subsequent characterization.

### 2.2 Characterization

The XRD patterns at ambient condition were acquired using a Bruker D2 Phaser diffractometer (Billerica, MA, USA) equipped with a Cu radiation (λ=1.54 Å) source and operated at 10 mA and 30 kV. The UV-Vis-NIR diffuse reflectance spectroscopy (DRS) were measured using Agilent Cary 5000 spectrophotometer with the use of $BaSO_4$ as a reflectance standard. The photoluminescence excitation (PLE) and photoluminescence (PL) spectra were recorded using a spectrofluorometer (Edinburgh

Instruments, FLS-1000) combined with a steady-state and fluorescence lifetime fluorescence spectrometer, equipped with a Hamamatsu R928 PMT-900 detector (185-900 nm) and NIR-PMT (up to 1700 nm). The excitation source of PL spectra was a 450 W Xe900 continuous xenon lamp, while for decay curves, a 60 W μF2 pulsed xenon microsecond flash lamp was used. Temperature dependent stead-state PL speatra and decay time measurements were performed using an ARS cryostat (model CS204AE-FMX-1AL) coupled with the FLS1000 spectrofluorometer and a Lake Shore 335 temperature controller. The persistent luminescence intensity was recorded after 15 min of exposure to UV or X-ray illumination. Then, TL glow curves were measured from 300 to 570 K by increasing the temperature at a heating rate of 1 K/s. The electroluminescence spectra of pc-NIR LED device were recorded using an OHSP-350M LED Fast-Scan Spectrophotometer, covering a wavelength range of 350-1050 nm (Hangzhou Hopoo Light&Color Technology Co.,Ltd. ).

## 3. Results and discussion

### 3.1 Structural identification

$Ca_3Ga_2Sn_3O_{12}$, a member of rich family of $A_3B_2C_3O_{12}$ garnet-type structure with $A^{2+}$ ions situated at $D_2$ sites, $C^{4+}$ at $S_4$ site, and $B^{3+}$ at $C_{3i}$ site (Figure 1a), crystallizes in the cubic ($O_h^{10}$-$Ia3d$ (230)) space group, as determined by X-ray diffraction at room temperature, where $GaO_6$ octahedra undergoes slightly trignal distortion with respect to ideal octahedral geometry [39].As depicted in Figure 1b, all observed peaks can be well indexed to the standard data (COD 1001563 [40]), confirming that desired compounds were readily prepared by solid-state reaction. The deliberate introduction of $Fe^{3+}$ ions into the octahedral $GaO_6$ sites does not seem to have any discernible impact on the crystal structure. As the concentration of $Fe^{3+}$ ions increases, the principal diffraction peaks are shifted towards lower angles. Structural parameters for CGSG:xFe (x=0, 0.02) samples were determined by the Rietveld refinement method with FullProf software, adopting a cubic structure with space group $O_h^{10}$-$Ia3d$ (230). The obtained parameters are listed in Table 1. The results demonstrate an increase in the cell parameter, in accordance with observed shifts of the main diffraction peaks. The increase in the cell parameter is attributed to the larger ionic radius of $Fe^{3+}$ ions (0.645 Å, CN = 6) compared to $Ga^{3+}$ ions (0.620 Å, CN = 6).

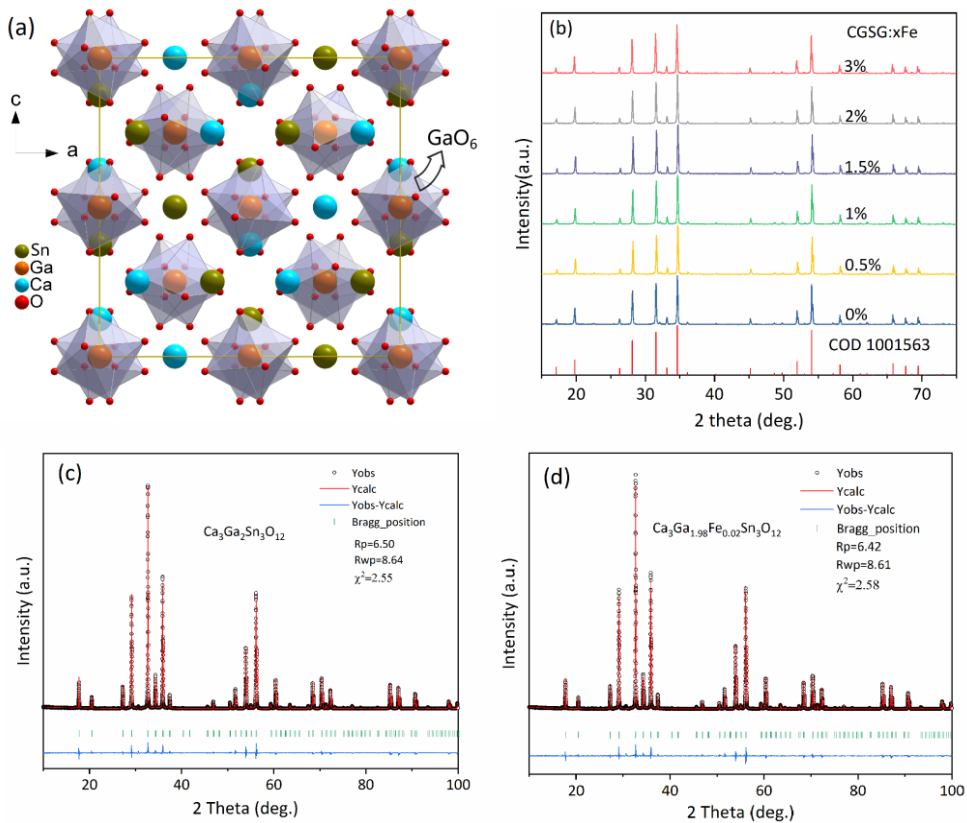

**Fig. 1.** (a) Schematic diagram of the crystal structures of $Ca_3Ga_2Sn_3O_{12}$ and $GaO_6$ octahedral coordination geometry. (b) XRD patterns of $Ca_3Ga_{2-x}Sn_3O_{12}:xFe^{3+}$ (x=0～0.03) phosphors. (c) and (d) The Rietveld refinement patterns of CGSG:xFe (x=0, 0.02) samples.

Table 1　Structural parameters and main Rietveld refinement details for CGSG and CGSG:0.02Fe phosphors

| Parameters | CGSG | CGSG:0.02Fe |
|---|---|---|
| $a$ (Å) | 12.25200(7) | 12.25433(7) |
| $V$ (Å$^3$) | 1839.167(18) | 1840.216(18) |
| $R_p$ | 6.50 | 6.42 |
| $R_{wp}$ | 8.64 | 8.61 |
| Chi squared $\chi^2$ | 2.55 | 2.58 |

## 3.2　Photoluminescence properties of CGSG:xFe phosphors

The photoluminescence properties and decay kinetics of CGSG:xFe were investigated to elucidate the electronic structure of $Fe^{3+}$ in CGSG. The energy levels of

3$d$ transition metal ions in crystal field are typically described using the well-known Tanabe-Sugano (T-S) diagram [41]. It should be noted that, for the $d^5$ system, both the tetrahedral and octahedral crystal field levels follow a similar order [42]. However, the key factors for substitutional occupation in solids are the valence, ionic radii and chemical properties of host and dopant ions. In the present case, the even valence and proximity of ionic radii strongly suggest a preference for $Fe^{3+}$ ions substituting $Ga^{3+}$ ions in the octahedral sites. The typical DRS, PL and PLE spectra of CGSG:xFe phosphors recorded at room temperature are presented in Fig. 2a. Upon UV excitation at 298 nm, the PL spectra display a broadband NIR emission extending from 650 to 1000 nm with a maximum at 754 nm and a full width at half maximum (FWHM) of 89 nm. This observed emission band is ascribed to internal $^4T_1(G) \rightarrow {}^6A_1(S)$ transition of $Fe^{3+}$ ions in octahedral sites. The assignment of the PLE features to octahedral crystal field levels of $Fe^{3+}$ is based on the observed positions of $^4E(^4D)$ and $^4E,^4A_1(4G)$ levels, which remain nearly independent of crystal field strength [43],[44]. The PLE spectrum, monitored at 754 nm, features a predominant band at 289 nm arising from charge transfer between $O^{2-} \rightarrow Fe^{3+}$. Additionally, there are four less intensive bands at 325, 390, 408, and 460 nm corresponding to optical transitions from $^6A_1(S)$ to $^4T_1(^4P)$, $^4E(^4D)$, $^4T_2(^4D)$ and $^4E/^4A_1(4G)$ states, respectively. The observation is consistent with theoretical predictions [45] and experimental data for $Fe^{3+}$ ions situated in an octahedral environment [46],[47],[48]. Two relatively weak bands (inset in Fig. 2a) around 580 and 678 nm can be ascribed to the absorption of $^4T_2(^4G)$ (likely mixed with $^2T_2$) and $^4T_1(^4G)$ states, which are usually not detectable due to the low oscillator strength of the NIR absorption. The Stokes-shift of $^4T_1$ state is determined to be 1487 cm$^{-1}$. A similar result is observed in DRS of CGSG:0.02Fe. The identified energy levels of $Fe^{3+}$ in CGSG are summarized and tabulated in Table 2.

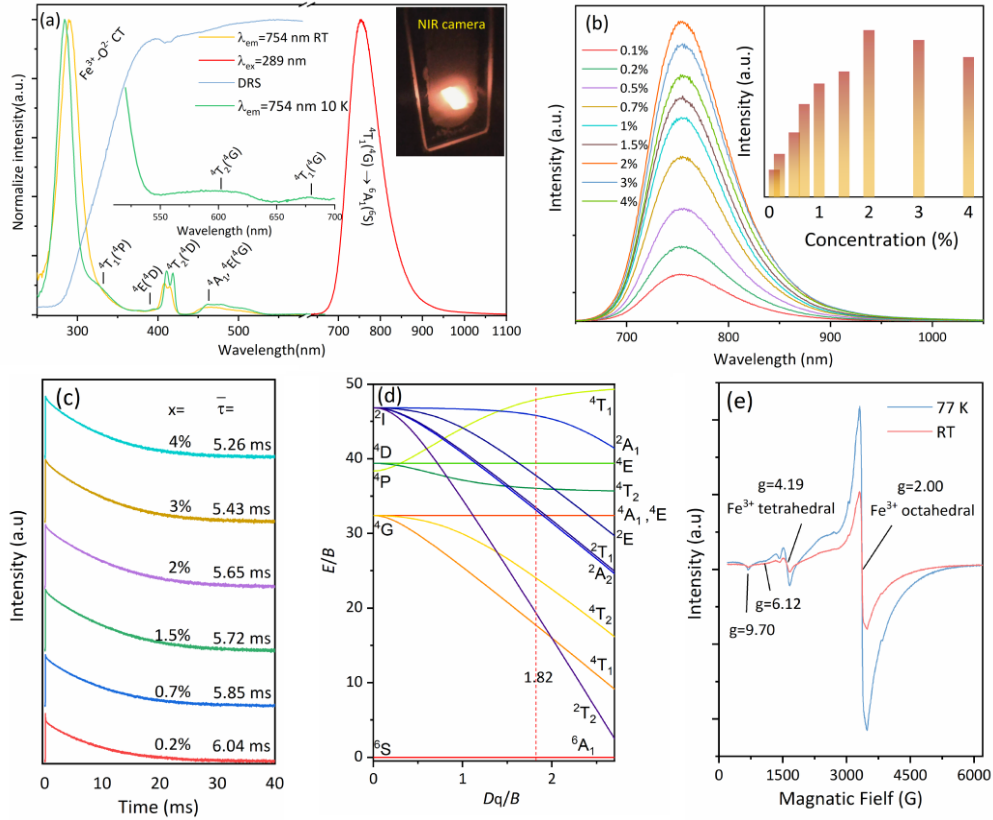

**Fig. 2.** (a) PLE (at RT and 10 K), DRS and PL (RT) spectra of CGSG:0.02Fe$^{3+}$ phosphor; Inset showing an enlarged figure of PLE spectrum. (b) PL spectra, and (c) decay curves of Ca$_3$Ga$_{2-x}$Sn$_3$O$_{12}$:xFe$^{3+}$ (x=0.001−0.04) phosphors; insert of (b) showing integrated PL intensity as function of Fe$^{3+}$ concentration. (d) The Tanabe-Sugano diagram for octahedral coordination of Fe$^{3+}$ ions. (e) EPR spectra of CGSG:0.02Fe$^{3+}$ measured at 77 K and RT.

**Table 2** Energy levels (all in cm$^{-1}$) and crystal-filed parameters for Fe$^{3+}$ ion in Ca$_3$Ga$_2$Sn$_3$O$_{12}$ (Ga octahedral site), obtained from experimental data and the fit of the crystal field theory

| Energy levels | Observed in PLE | Calculated from the fit of T-S matrices |
|---|---|---|
| $^4T_1$ ($^4P$) | 30718 (325 nm) | 31348 (319 nm) |
| $^4E$ ($^4D$) | 25641 (390 nm) | 26598 (376 nm) |
| $^4T_2$ ($^4D$) | 24510 (408 nm) | 24631 (406 nm) |
| $^4E,^4A_1$ ($^4G$) | 21739 (460 nm) | 21930 (456 nm) |
| $^4T_2$ ($^4G$) | 17241 (580 nm) | 18215 (549 nm) |
| $^4T_1$ ($^4G$) | 14749 (678 nm) | 14793 (676 nm) |

The concentration dependent PL spectra of CGSG:xFe (x=0.001-0.04) with the inset of their corresponding integrated PL intensity (Fig. 2b) indicate an optimal concentration at 2%. The shape and the maximum position of the spectrum remain nearly unchanged with varying Fe$^{3+}$ concentration. The luminescence decay profiles in Fig. 2c exhibit a multi-exponential decay feature and can be well fitted using a sum of two-exponential function: $I(t) = \sum_{i=1}^{2} A_i \exp(-t/\tau_i)$. The average decay lifetime $\bar{\tau}$ then was calculated using $\bar{\tau} = \frac{\sum_{i=1}^{2} a_i \tau_i^2}{\sum_{i=1}^{2} a_i \tau_i}$ equation, and it varies from 6.04 to 5.26 ms with increasing of $Fe^{3+}$ concentration. The concentration quenching is attributed to energy migration process between the adjacent $Fe^{3+}$ ions thought multipole-multipole processes or exchange interaction.

The crystal field strength $Dq$ indicates the symmetry of impurity ion site, whereas Racah parameters $B$ and $C$, associated with the admixture orbitals of $Fe^{3+}$ with its ligand $O^{2-}$, contribute to breaking selections rules and enabling electronic transitions. These parameters can be determined by analyzing energetic positions in PLE spectra and the T-S matrices. According to T-S theory [41], the energies of $^4A_1$, $^4E$ and $^4T_1$ bands are given in terms of Racah parameters $B$, $C$ and $Dq$ by the following expressions:

$$^6A_1(^6S) \rightarrow {}^4A_1, {}^4E(^4G) = 10B + 5C \tag{1}$$

$$^6A_1(^6S) \rightarrow {}^4E(^4D) = 17B + 5C \tag{2}$$

The energies of the $^4T$ states are determined by the solving of the third-order nonlinear equations and subsequently expressed in relation to Racah and crystal-field strength $Dq$ parameters. The best fit of the crystal-field theory yields the most probable values for $Dq$, $B$ and $C$ as follows: $D_q$=941 cm$^{-1}$, $B$=670 cm$^{-1}$, $C$=3047 cm$^{-1}$, $Dq/B$=1.41, $C/B$=4.55, which are comparable to those reported for octahedrally coordinated $Fe^{3+}$ in spinel $AB_2O_4$($A$=Mg, Zn;$B$=Al, Ga) [49],[50], $LiGa_5O_8$ and $LiGaO_2$ [51] compounds. Theoretical values of the free-ion Racah parameters are much higher, such as those reported in ref. [52], where $B_0$ = 1130 cm$^{-1}$ and $C_0$ = 4111 cm$^{-1}$, and ref. [53] calculated values at 1296 and 4826 cm$^{-1}$, respectively. The experimental values differ from those derived theoretically and are determined to be $B_0$=670 cm$^{-1}$, $C_0$=4932 cm$^{-1}$ [54]. The reduction of the Racah parameters in the host environment is a result of the covalence effect, which is quantified by the so-called nephelauxetic parameter $\beta_1$($\beta_1 = \sqrt{(B/B_0)^2 + (C/C_0)^2}$) [55]. The calculated $\beta_1$ parameter is 1.03, indicating a predominant covalent character of the $Fe^{3+}$-$O^{2-}$ bond rather than an ionic one. Such bond can induce a significant orbital mixing between $Fe^{3+}$ and ligand ions. Consequently, the Laporte selection rule is largely lifted, enabling electronic transitions to occur for octahedral $Fe^{3+}$ ions. On the other hand, the strong luminescence of $Fe^{3+}$ can in part be attributed to its occupation of $GaO_6$ octahedral sites characterized by very low symmetry (point group $C_{3i}$). This lowered symmetry further gives rise to high possibility of electronic transitions, thereby contributing to the enhancement of luminescence in CGSG:Fe phosphor.

Room-temperature and low-temperature (77 K) EPR spectra of CGSG:0.02$Fe^{3+}$ were recorded as a part of our assessment as for site occupation of $Fe^{3+}$ in the CGSG lattice. The $Fe^{3+}$ ion adopts five unpaired electrons in the *d* orbital with S=5/2 and yields five-line spectrum corresponding to S-state transitions: 5/2↔3/2, 3/2↔1/2, 1/2↔-1/2, -1/2↔-3/2, and -3/2↔-5/2. However, the 1/2↔-1/2 transition is only observed in present study due to the large crystal field splitting. As shown in Fig. 2e, the typical EPR spectrum is characterized with an intense resonance signal at g≈2.00, a weak signal at g≈4.19, with a shoulder around g≈6.12, indicating different coordination environments of $Fe^{3+}$ ions. The weak signal around g≈4.19 can be attributed to high spin $Fe^{3+}$ ions in tetrahedral crystal field, while the intense signal with g≈2.00 is readily assigned to $Fe^{3+}$ ions in an octahedral environment [56],[57]. The additional signals with higher g values and relatively weak intensity may originate from tetrahedral associated oxygen vacancy, where a defect complex of the type $Fe^{3+}$-Vo could be formed [58]. Based on the EPR analysis, it can be concluded that the photoluminescence emission of CGSG:Fe stems from $Fe^{3+}$ ions in $GaO_6$ octahedral sites.

**3.3 Temperature dependence of PL and PL decay time**

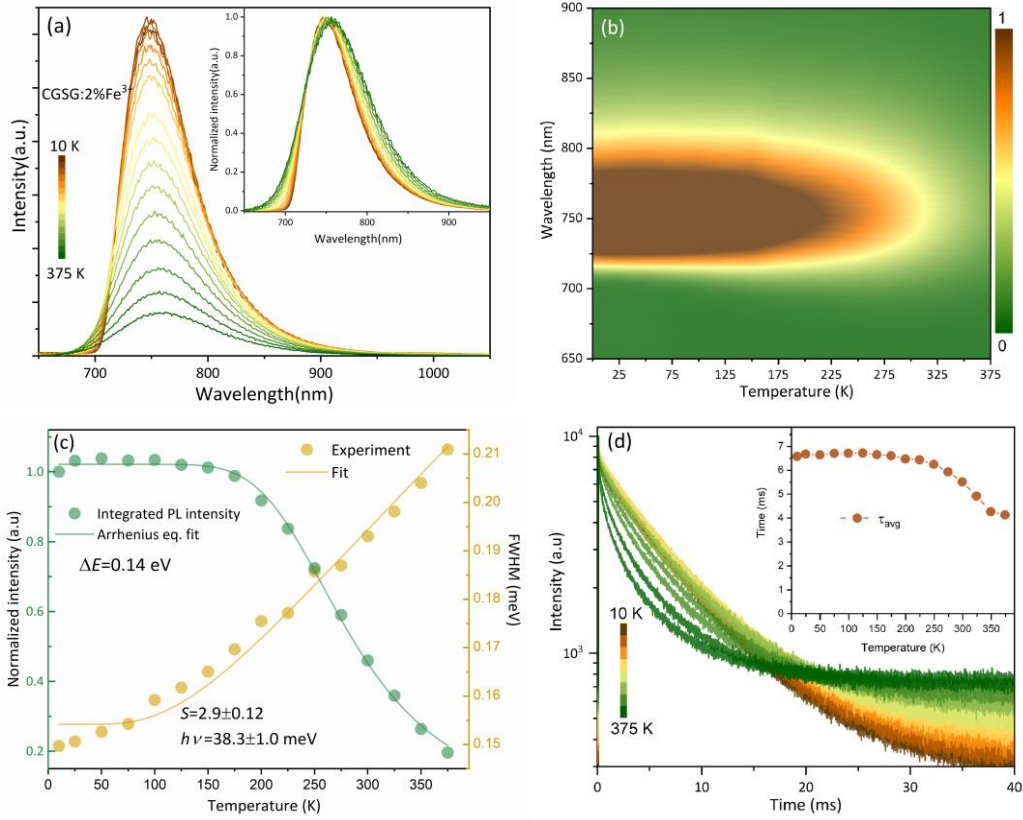

**Fig. 3.** (a) Temperature-dependent PL spectra and (b) 2D temperature-dependent PL spectra for CGSG:0.02Fe$^{3+}$ phosphor in the 10-375 K range. (c) The integrated PL intensity and FWHM as a function of temperature. (d) temperature dependent decay curves of 754 nm emission under 298 nm excitation, the insert showing the shorter, longer and average decay lifetime as function of temperature.

The temperature-dependent PL measurements of CGSG:0.02Fe phosphor were carried out to evaluate its thermal stability, as depicted in Fig. 3a and 3b. The PL spectrum exhibits a slight red shift and the integrated PL intensity decreases as the temperature rises, as shown in Fig. 3c. Similar trend was observed in PL decay curves (Fig. 3d), the average decay lifetime of Fe$^{3+}$ luminescence decreases with increasing temperature. The elevated tail in the decay curves at high temperatures might foreshadow afterglow properties. With increasing temperature, the maxima of $^4T_2 \rightarrow {}^6A_1$ band showed a slight shift toward longer wavelengths starting with 748 nm at 10 K to 760 nm at 375 K. The observed red shift in PL is primarily attributed to increased electron-phonon interaction at high temperatures. The integrated PL intensities as a function of temperature can be well fitted using the following Arrhenius formula:

$$I(T) = \frac{I_0}{1 + Ae^{\Delta E/k_B T}} \qquad (3)$$

where $I(T)$ and $I_0$ are the integrated PL intensities at $T$ and at the lowest temperature,

respectively, $\Delta E$ is the activation energy, and $k_B$ is the Botzmann constant. $\Delta E$, representing the energy separation between minima of $^4T_1$ state and its crossover point to $^6A_1$ ground state, is calculated to be $\Delta E$=140 meV, suggesting a stronger electron-phonon coupling (EPC) of the $^4T_1 \rightarrow {}^6A_1$ emission of $Fe^{3+}$ in CGSG. The strength of EPC can be estimated by by analyzing the vibration of FWHM as a function of temperatures, as depicted in Fig. 3c. This relationship is described by the Boltzmann distribution function given by[21]

$$FWHM = 2.36 \times h\nu \times \sqrt{S} \times \sqrt{\coth\left(\frac{h\nu}{2kT}\right)} \quad (4)$$

where $h\nu$ is the effective phonon energy that interacts with the electronic transitions, $T$ is the Kelvin temperature, $S$ is the dimensionless Huang-Rhys parameter and $k$ is the Boltzmann constant ($0.8617 \times 10^{-5}$ eV). The obtained values of $S$ and $h\nu$ are 2.9 and 38.2 meV. Therefore, the strong thermal quenching of CGSG:Fe is attributed to the strong electron-phonon coupling effect. The Stokes-shift can be estimated using $E_{Stokes-shift}=(2S-1)h\nu$, producing a value of 1479 cm$^{-1}$, in a good agreement with experimental data (1487 cm$^{-1}$).

**3.4 NIR persistent luminescence properties of CGSG:Fe**

In addition to its steady-state photoluminescence, CGSG:Fe exhibits efficient NIR persistent luminescence (PersL), making it a multifunctional material. The PerL decay curves monitored at 754 nm were recorded after irradiation by UV light or X-ray for 15 min. As shown in Fig. 4a-b, the persistent duration over 1 h can be traced. The PersL spectra acquired after the removal of the irradiation were also presented in Fig. 4c, showing similar shape with steady-state PL spectrum. Notably, the more intense NIR PersL with longer time can be induced by X-ray irradiation. Such luminescent materials featuring long NIR PersL properties hold significant potential for background-free in vivo biomedical multi-cycle imaging when subjected to safe X-ray radiation doses if they are synthesized in nano size in the future. The measurement of thermoluminescence (TL) curve is widely considered as one of the most powerful techniques for gaining precise insights into the traps, including trap distribution and trap depth [59]. The TL curve monitoring at 754 nm for CGSG:0.02Fe, presented in Fig. 4d, demonstrates a maximum of TL intensity around 415 K when the temperature was raised with a heating rate of 1K s$^{-1}$. The only one broad band in TL curve implies a trap system likely comprised of single discrete energy level with simple first order kinetics. The trap depth can be roughly estimated by $E(eV)=T_m(K)/500$ [60], here $T_m$ is the temperature at the maximum of TL glow. Thus, trap depth is determined to be 0.83 eV. Furthermore, the optically stimulated luminescence (OSL) spectrum excited by 808 nm laser further confirms the existence of electron traps.

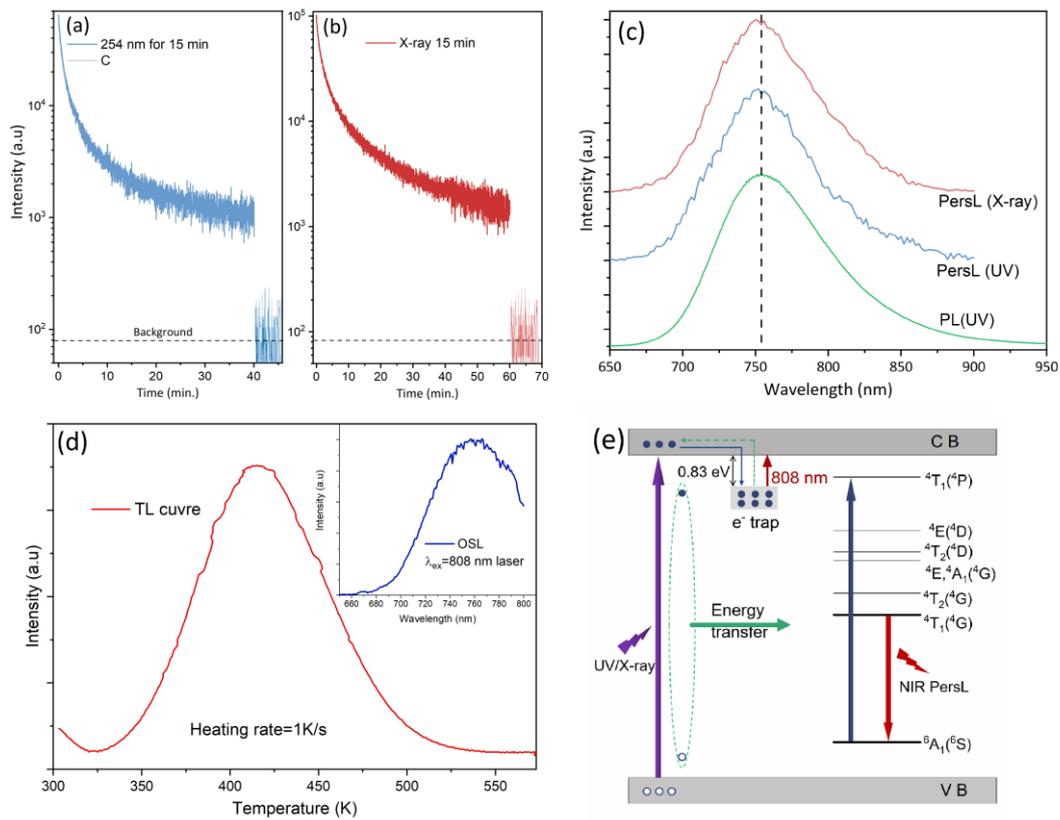

**Fig. 4.** NIR persistent luminescence decay curve monitored at 754 nm after irradiation by (a) 298 nm light and (b) X-ray for 15 min. (c) The normalized PL and PersL spectra of CGSG:0.02Fe phosphor. (d) TL curve monitoring at 754 nm with a time interval of 1 min after removal irradiation of 254 nm light for 15 min over the range of 298-570 K; the insert showing OSL spectrum excited by 808 nm. (e) UV light or X-Ray induced NIR PerL mechanism of the CGSG:0.02Fe phosphor.

Based on the above analysis, we propose a preliminary physical model to explain the observed NIR PersL in CGSG:Fe phosphors (Fig. 4e). In this model, the traps are considered to be electron traps located below the CB minimum of CGSG about 0.83 eV, likely originating from oxygen vacancies or other intrinsic defects. Upon UV and X-ray excitation, the electrons are excited from the valance band (VB) to the conduction band (CB), generating holes in the VB. The electrons and holes can either recombine or be captured by electron and hole traps in the CGSG, respectively. Some of the excitation energy is transferred to $Fe^{3+}$ ions, resulting in the PL of CGSG:Fe. Meanwhile, the excited electrons in the CB are captured by the electron traps below the CB. After the cessation of illumination, the captured electrons in the trap are gradually released to the CB *via* thermal stimulation or photo stimulation, leading to non-radiative recombination with the holes and transferring the recombination energy to $Fe^{3+}$ ions, which generates efficient NIR PersL.

**3.5 Fabrication of NIR pc-LED and its potential applications in night vision**

A broadband NIR pc-LED was fabricated by coating CGSG:0.02$Fe^{3+}$ phosphor on a commercial 310 nm LED chip. When the NIR pc-LED is lit up, the images of the NIR pc-LED are taken by a visible camera and NIR camera, respectively, as shown in the inset of Fig. 5a. The electroluminescence (EL) spectra exhibit an efficient broad NIR emission band in the 660-950 nm attributed to CGSG:0.02$Fe^{3+}$ phosphor, demonstrating an increase as the driven current is raised. However, the achievement of a substantially high NIR output power was impeded by the threshold current of the UV chip, resulting in the pc-LED reaching an NIR output power of 0.6 mW under a forward current of 40 mA. The potential application as a compact NIR light source for night vision was examined, as shown in Fig. 5b-g. The pictures were captured with different cameras under the irradiation of w-LED or packaged NIR pc-LED. After turning off the w-LED, no objects could be observed by the camera. In contrast, when the encapsulated NIR pc-LED was lit, clear images of a toy and a QR code can be observed using the NIR camera. These results demonstrate the potential application of the CGSG:$Fe^{3+}$ phosphor in the realm of night vision.

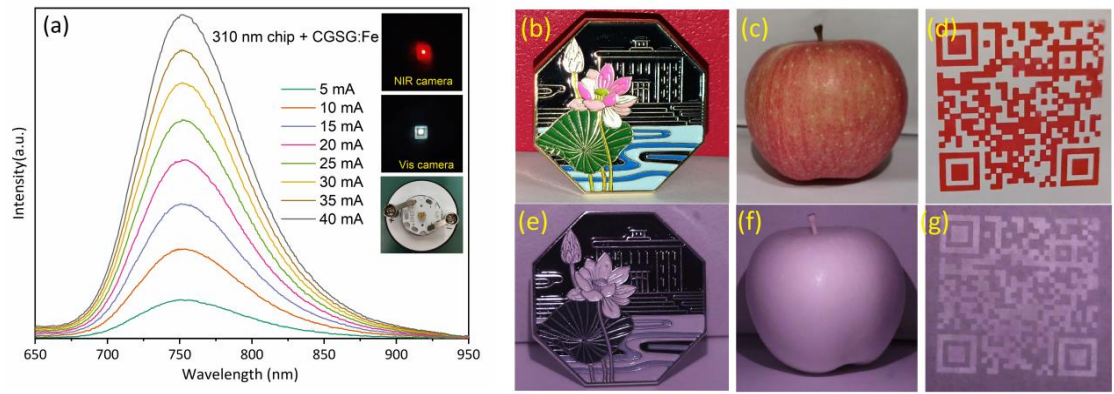

**Fig. 5.** (a) PL spectra of fabricated pc-NIR LED coated with CGSG:0.02$Fe^{3+}$ phosphor with increase of driven current; insert showing the device and photos taken with visible and NIR camera. The images of the objects were captured under the following conditions: (b) - (d) using visible camera with w-LED light on, (e) - (g) using NIR camera with NIR pc-LED light on.

## 4. Conclusion

In conclusion, a series of $Fe^{3+}$-activated broadband garnet-type $Ca_3Ga_2Sn_3O_{12}$ phosphors was designed and successfully synthesized. Under 298 nm excitation, the phosphor exhibits a broadband NIR emission in the 650-1000 nm range, peaking at 754 nm with a FWHM of 89 nm, attributed to $^4T_1(G) \rightarrow {}^6A_1(S)$ transition of octahedrlly coordinated $Fe^{3+}$ ion. The crystal-field and Racah parameters of $Fe^{3+}$ were evaluated based on experimental data and Tanabe-Sugano crystal-field theory. Furthermore, the NIR persistent luminescence lasting over 1 h was detected following UV or X-ray irradiation. The observed NIR persistent luminescence is proposed to involve a mechanism related to electron traps. Furthermore, a NIR pc-LED was fabricated by coating synthesized phosphor on a UV chip, and its performance was evaluated to assess its potential suitability as a NIR light source. Our discovery of new type of UV and X-ray activated broadband NIR luminescence phosphors with efficient NIR PerL paves the way for developing alternative $Cr^{3+}$-free multifunctional NIR luminescence materials, thereby broadens the scope of potential applications in diverse fields.

## Acknowledgements

This work was financially supported by National Natural Science Foundation of China (12004062), the Science and Technology Research Program of Chongqing

Municipal Education Commission (KJZD-K202300612, KJQN202100615, KJQN202100640, KJZD-M20230060, KJQN202300652), and the Natural Science Foundation of Chongqing (cstc2021jcyj-msxmX0277, CSTB2022NSCQ-MSX0366, sl202100000301). Andrzej thanks to the financial support from Polish National Science Center program SHENG2 of Poland-China cooperation (2021/40/Q/ST5/00336).